\documentclass{aa}
\usepackage{txfonts}
\usepackage{natbib}
\usepackage{graphicx}
\bibpunct{(}{)}{;}{a}{}{,} 
\usepackage{longtable}
\usepackage{xspace}

\begin{document}
\title{A photometric search for active Main Belt asteroids}

\author{S. Cikota \inst{1}
\and J. L. Ortiz\inst{2}
\and A. Cikota \inst{3}
\and N. Morales \inst{2}
\and G. Tancredi \inst{4}}

\institute{Physics Department, University of Split, Nikole Tesle 12, 21000 Split, Croatia. \email{scikota@pmfst.hr}
\and Instituto de Astrof\'{\i}sica de Andaluc\'{\i}a - CSIC, Apt 3004, 18008 Granada, Spain.
\and Institute for Astro- and Particle Physics, University of Innsbruck, Technikerstr. 25/8, A-6020 Innsbruck, Austria.
\and Observatorio Astronómico Los Molinos DICYT-MEC Cno. de los Molinos 5769, 12400 Montevideo, Uruguay.}

\titlerunning{MBCsearch}

\abstract{It is well known that some Main Belt asteroids show comet-like features. A representative example is the first known Main Belt comet 133P/(7968) Elst-Pizarro. If the mechanisms causing this activity are too weak to develop visually evident comae or tails, the objects stay unnoticed. 
We are presenting a novel way to search for active asteroids, based on looking for objects with deviations from their expected brightnesses in a database. Just by using the MPCAT-OBS Observation Archive we have found five new candidate objects that possibly show a type of comet-like activity, and the already known Main Belt comet 133P/(7968) Elst-Pizarro.
Four of the new candidates, (315) Constantia, (1026) Ingrid, (3646) Aduatiques, and (24684) 1990 EU4, show brightness deviations independent of the object's heliocentric distance, while (35101) 1991 PL16 shows deviations dependent on its heliocentric distance, which could be an indication of a thermal triggered mechanism.
The method could be implemented in future sky survey programmes to detect outbursts on Main Belt objects almost simultaneously with their occurrence.}


\maketitle

%

\section{Introduction}

Main Belt comets are a newly recognized group of objects originating in the main asteroid belt \citep{HandJ2006} that show sporadic comet-like appearance. An already well-known textbook example of the Main Belt comets is the object 133P/(7968) Elst-Pizarro. It was reported for the first time in 1979 as a minor planet, being completely stellar in appearance. In 1996 Eric Elst and Guido Pizarro found its cometary aspect when the object was near perihelion \citep{Elst1996}. Subsequently, the cometary-like activity appeared again around the perihelion at the end of 2001 \citep{Hsieh2004}. Because of the persistent appearance of the comet-like characteristics, the object was classified as a comet and today it is one of just a few objects that have a dual status, that is they have both minor planet and comet designations.

Since the discovery of comet-like activity on 133P/(7968) Elst-Pizarro, just a handful of similar Main Belt objects have been found. \cite{Jewitt2012} proposed possible mechanisms for producing mass loss from asteroids, but the cause of the activity of all known Main Belt comets is still unknown and the cometary nature of these objects is still under debate.

Statistically, there are indications that there are many more Main Belt comets than the known ones. Based on the discovery of one active object among 599 observed ones, now known as 176P/LINEAR, \cite{Hsieh2009} has suggested that there could exist $ \sim $100 currently active Main Belt comets among low inclination, kilometer-scale outer belt objects.

A promising statistical indication of many more Main Belt comets are estimations of impact rates in the Main Belt. The recent outbursts on P/2010 A2 (LINEAR) \citep[e.g.][]{MorenoA2, Snodgrass2010} and (596) Scheila \citep[e.g.][]{Moreno596, Jewitt2011} have proven that impacts can also be causes for the comet-like appearance of Main Belt asteroids. For example, the impactor size of P/2010 A2 (LINEAR) is estimated to a body of diameter 6-9 m \citep{Jewitt2010, Snodgrass2010, Larson2010}. The roughly estimated impact rate of impacts of this size is every 1.1 billion years for a parent body of diameter 120 m, which corresponds to one impact every $ \sim $12 years somewhere in the asteroid belt \citep{Snodgrass2010, Bottke2005}. By reducing the impactor sizes the impact rate increases and it is not yet known what effects can cause much smaller impactors in the range of $\sim$0.1-1 m, which probably are not negligible.

All the Main Belt asteroids that were classified as comets showed a diffuse, instead of a star-like, visual appearance. 
Hunting for comet-like Main Belt objects by searching for typical cometary features, like tails or comae, requires a lot of telescope time in middle and large class telescopes and does not guarantee success in detecting them all.

Although a search for new Main Belt comet candidates performed by \cite{Sonnett2011}, searching in the Thousand Asteroid Light Curve Survey (TALCS) images \citep{Masiero2009}, resulted in no new candidates, evidence was found that $\sim$5$\%$ of Main Belt asteroids might be active at low levels, in form of a faint tail which cannot be detected individually, but collectively in the TALCS data set.

A mechanism of mass loss that might offer an explanation of weak activities on small bodies is seismic shaking \citep{Tancredi2012} induced by the release of energy by the liberation of internal stresses, the reaccommodation of material, or thermal cracking. Small impacts might generate shock waves that can propagate to the body interior and globally shake the object. 
Simulations of surface shaking in low-gravity environments like those of small solar system bodies have shown that particles can be ejected from the surface at very low relative velocities \citep{Tancredietal2012}.
Periodic recurrence of the activity on some objects can be explained by meteoroid streams in the Main Belt, causing many of small impacts and triggering some of the mass-loss mechanisms.

A mass-loss mechanism possibly related to asteroid collisions is rotational instability \citep[e.g.][]{Jewitt2012}. In addition, the Yarkovsky–O'Keefe–Radzievskii–Paddack (YORP) \citep{Bottke2006} effect tends to modify the objects' spin rates. Close encounters with planets \citep{Scheeres2004} or large asteroids can also affect the objects' spin rates. Reaching the critical rotation frequency by increasing the spin rates can lead to the deformation and rotational fission of objects \citep{Comito2011, Rossi2012}, or just to the ejection of dust grains from the objects' surface.

It is difficult to guess how much material is produced by the suggested mass-loss mechanisms, but we believe that there must exist objects with some weak type of cometary activity, producing only a thin coma which is barely detectable by the visual examination of images. For example, this could include activity triggered by small impactors or be caused by electrostatic levitation \citep{Jewitt2012}.

In this paper we present a search for objects that indicate some type of weak activity, in form of deviations from their expected brightness. With a view to examining as many objects as possible, our search for new comet-like Main Belt objects is based on the data available in the MPCAT-OBS Observation Archive. Even though the photometric accuracy of the MPCAT-OBS Observation Archive is poor, it still has some potential, especially for large objects that become relatively bright near their opposition points, ensuring high signal-to-noise ratios and good photometric accuracy using today's sensitive instruments.

\section{Methods and materials}

The main idea was to compare every object's observed brightness with its expected brightness.
With the goal to have as many observations per object as possible over a minimum of three oppositions, we decided to use the MPCAT-OBS Observation Archive covering only numbered objects. The database used from October 2011 contains $ \sim $75 million observations, covering $ \sim $300\,000 numbered objects.

To avoid the overlapping of measurements collected through various photometric bands, it was important to choose observations collected by using just one photometric band. Because a lot of sky surveys observed in the visual (V) photometric band, and for them we can expect that their data are relatively precise, but also because of tests that showed that the observations collected in the V band are in good agreement with the Minor Planet Center or JPL Solar System Dynamics database estimated magnitudes, it was decided to use observations collected in the V photometric band.
The computations of expected brightnesses for every observation from the MPCAT-OBS Observation Archive are based on the H-G magnitude system \citep{Bowell1989}, and were calculated by using the Astronomical Ephemeris library PyEphem for Python \citep{Rhodes2011}. The orbital and physical parameters of the objects, which were necessary to compute the expected brightnesses, were taken from the Minor Planet Center Orbit Database (MPCORB), version prepared on October 28, 2011. Additionally, a few examples of the computed brightnesses have been compared to the computations using JPL HORIZONS, to verify they show good correspondence.

\begin{figure}
\centering
\includegraphics[width=88mm]{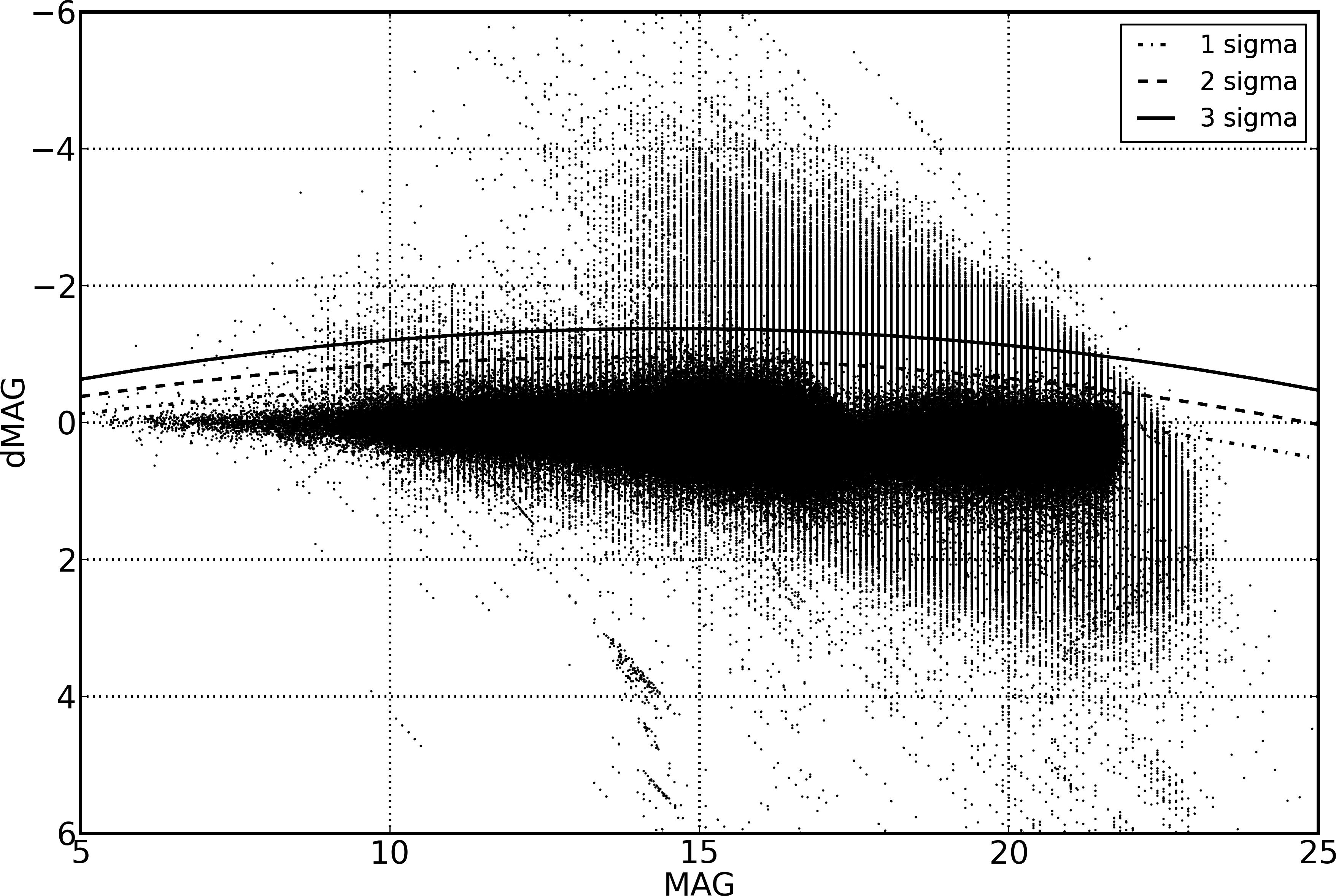}
\caption{Brightness deviations (dMAG) of all measurements in the visual photometric band taken from the MPCAT-OBS Observation Archive, shown versus their observed brightness (MAG). To filter out the objects we were interested in, we considered only measurements above the full, second-order curve, representing the 3-sigma limit.}
\label{fig1}
\end{figure}

To visualize the brightness deviations between the expected and observed brightness, the data was shown in a dMAG versus MAG plot (Fig.~\ref{fig1}), where MAG (x-axis) indicates the observed brightness and dMAG (y-axis) the difference between the observed and expected brightness, defined as in Eq. 1. The plot includes about 24 million brightness measurements in the V photometric band from all $ \sim $300\,000 objects contained in the MPCAT-OBS Observation Archive.
$$ dMAG = observed\,brightness - expected\,brightness \eqno(1)$$

To filter out the objects we were interested in, we considered only measurements of objects with negative brightness deviations (brightness increasing) greater than 3-sigma for a minimum of five measurements per object. The sigma values were calculated in one magnitude steps for all V measurements from magnitude 8.0 to 23.0. Through the calculated sigma values a second-order polynomial trend-line was fitted. In Fig.~\ref{fig1} trend-lines for the 1-sigma, 2-sigma, and 3-sigma limits are shown. The 3-sigma condition that we used to filter the data is defined as
$$ dMAG < 0.0082 \cdot MAG ^{2} - 0.2381 \cdot MAG + 0.3549. \eqno(2)$$
In this way, $\sim$1700 object designations were extracted.

For each of the $\sim$1700 extracted objects, we generated plots like those shown in Figs.~\ref{fig8}-\ref{fig3}, and \ref{fig4}-\ref{fig7}. The upper plots show the brightness deviations (y-axis) versus time given in Julian Date (x-axis). The measurements, shown in various markers for observations collected by different observatories (labelled with observatory codes assigned by the Minor Planet Center), are bunched into small groups that correspond to observations collected in different oppositions. The lower plots show the brightness deviations (y-axis) versus heliocentric distance given in astronomical units (x-axis). The heliocentric distance for every observation was also computed in Python, by using the Astronomical Ephemeris library PyEphem \citep{Rhodes2011}.
The object candidates showing unusual brightness deviations were extracted by visual examination of the plots.

\section{Results and Discussion}

The main principle of the visual examination of the plots was to search for objects that show brightness deviations at least over a few nights. Large brightness deviations that include just a few measurements over one night, reported by just one Observatory code, are mostly produced by close encounters of objects with bright stars and their blooming spikes in the images.

\begin{figure}
\centering
\includegraphics[width=8cm]{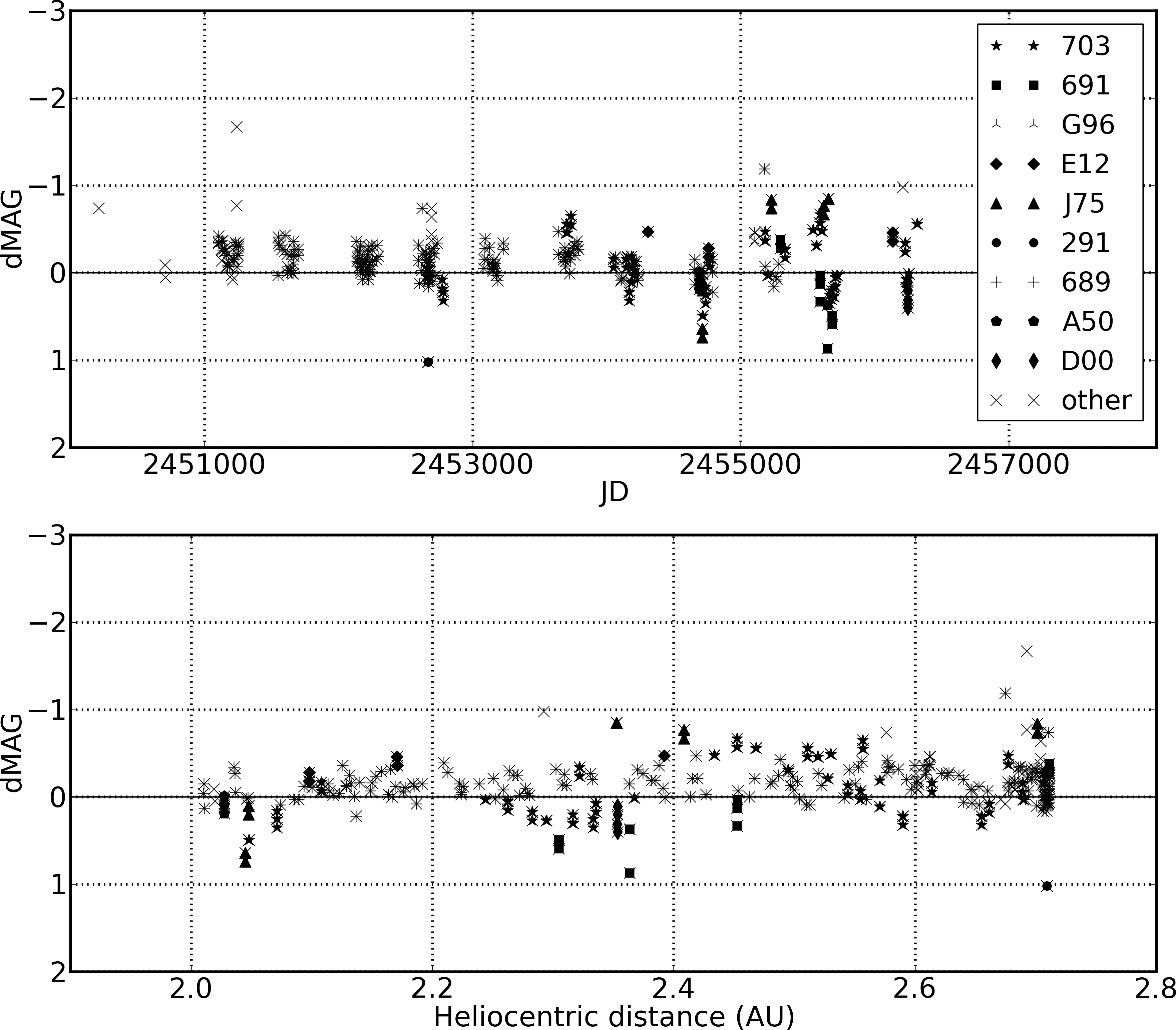}
\caption{Data plots for (306) Unitas containing 390 points. The upper plot shows the brightness deviations versus time given in Julian Date. The lower plot shows the brightness deviations versus heliocentric distance given in astronomical units. The various markers represent the different observatory codes assigned by the Minor Planet Center.}
\label{fig8}
\end{figure}

\begin{figure}
\centering
\includegraphics[width=8cm]{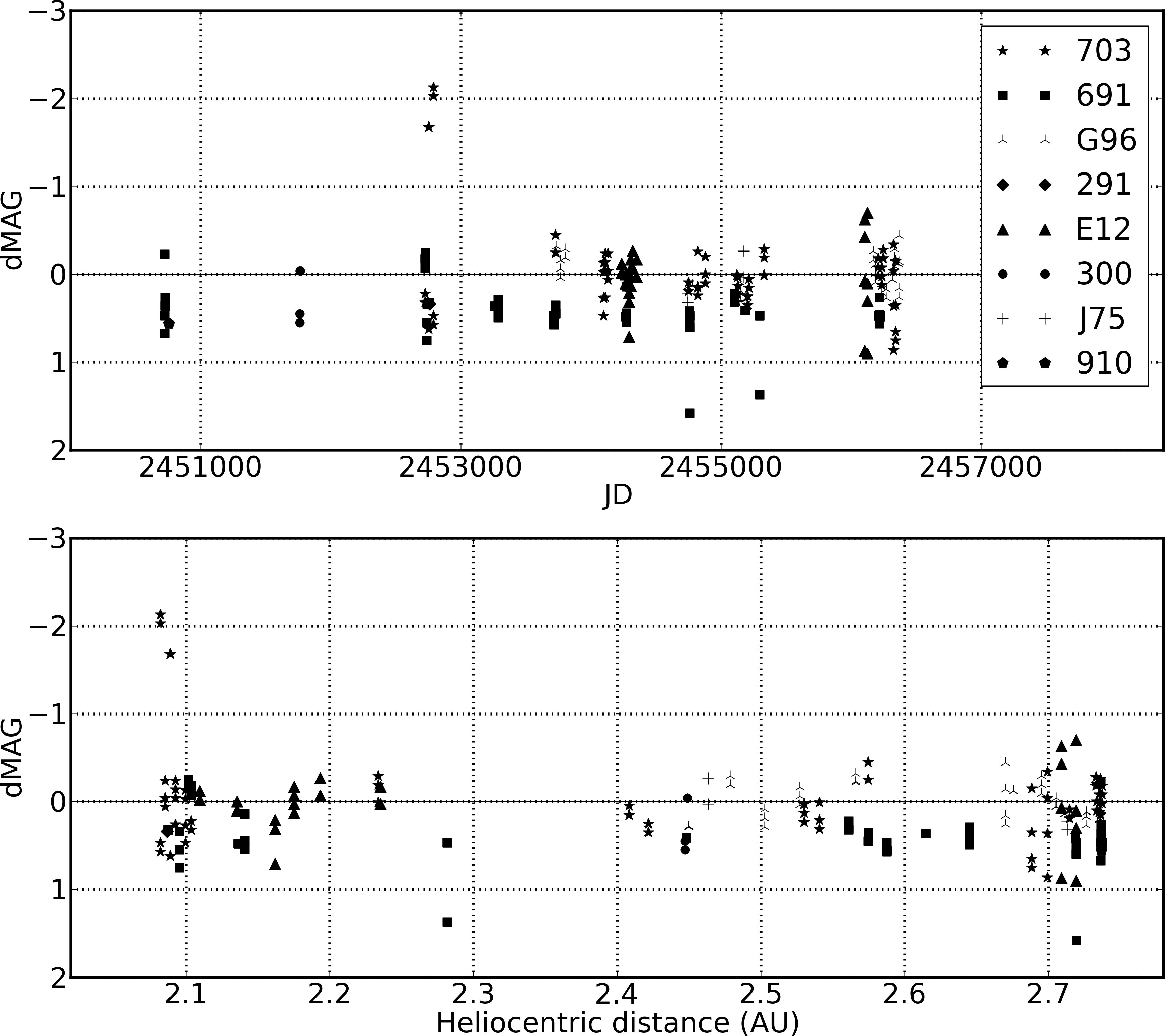}
\caption{Data plots for (10059) 1988 FS2 containing 251 points. The upper plot shows the brightness deviations versus time given in Julian Date. The lower plot shows the brightness deviations versus heliocentric distance given in astronomical units. The short brightness increase at JD$ \sim $2452750, containing just a few measurements of observatory code 703, is caused by some image artefacts. These short outbursts are responsible for most of the false detections found while using our filtering requirements. The various markers represent the different observatory codes assigned by the Minor Planet Center.}
\label{fig9}
\end{figure}

In this way, we extracted six objects which show unusual long-term brightness deviations, lasting within their whole oppositions, and changing their intensity with time. Some of the objects show a possible correlation of their brightness deviation versus heliocentric distance, while some of them do not. One of the extracted objects was the already known Main Belt comet 133P/(7968) Elst-Pizarro, while five of them were objects without previously observed signs of cometary activity. The maximum observed apparent magnitudes of the candidates ranges between 13.9 mag and 16.7 mag, presuming good photometric accuracy, at least for observations collected near the opposition points. Almost all of the brightness deviations were detected by multiple observatories at the same time, confirming the reliability of the deviations. To demonstrate data plots of objects that do not indicate any activity in the form of brightness deviations, we have randomly chosen (306) Unitas (Fig.~\ref{fig8}) and (10059) 1988 FS2 (Fig.~\ref{fig9}) from the 1700 extracted objects by using our filtering requirements.

We note here that in case of an inaccurate estimation of the objects' absolute magnitudes, the brightness deviations would have a constant offset for any time, as high as the aberration of the real absolute magnitude, but would not imply any activity. 
The new object candidates and their basic orbital elements are listed in Table~\ref{tb1}.

\begin{figure}
\centering
\includegraphics[width=8cm]{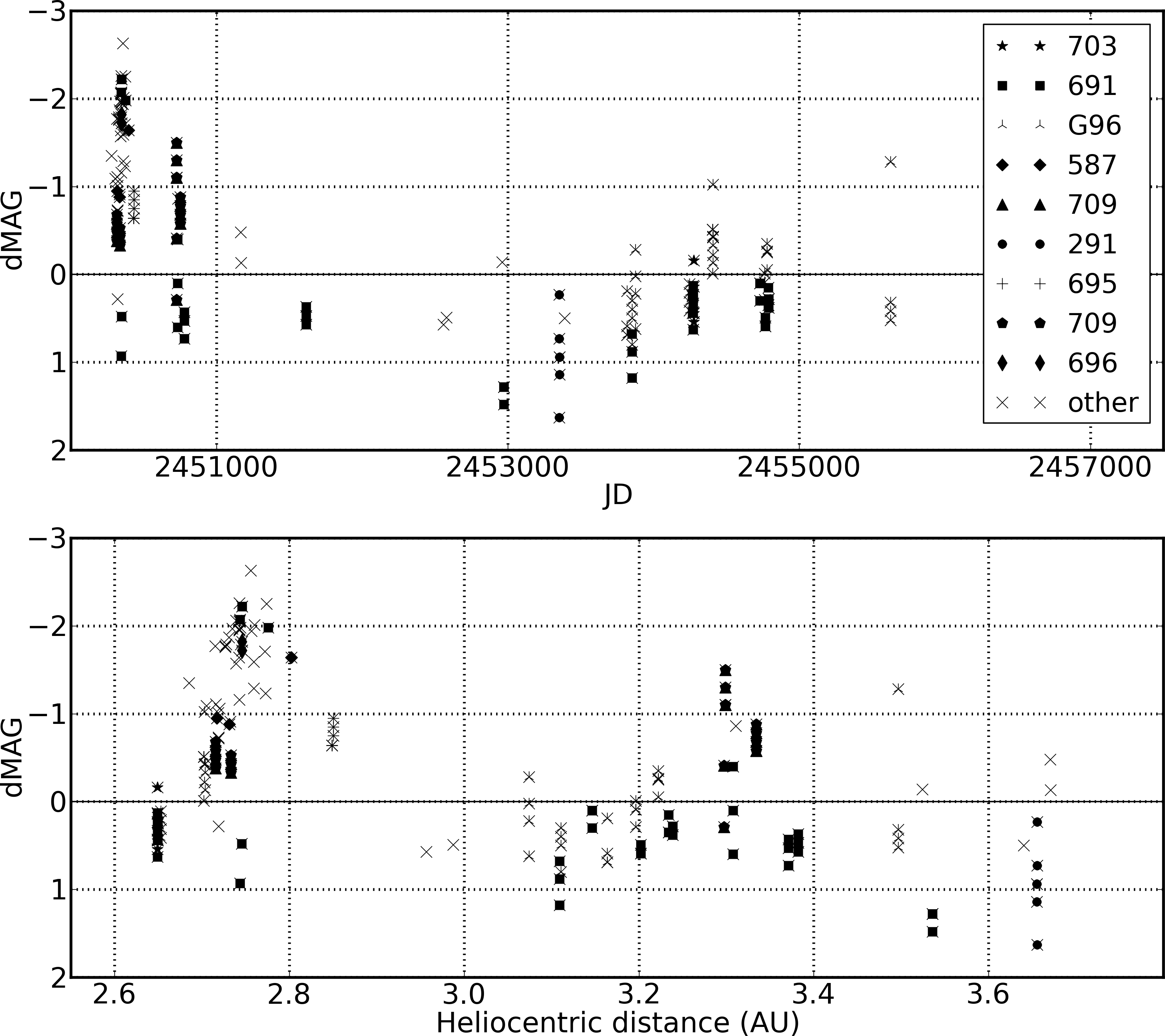}
\caption{Data plots for 133P/(7968) Elst-Pizarro containing 169 points. The upper plot shows the brightness deviations versus time given in Julian Date. The lower plot shows the brightness deviations versus heliocentric distance given in astronomical units. The various markers represent the different observatory codes assigned by the Minor Planet Center.}
\label{fig2}
\end{figure}

The data plots of 133P/(7968) Elst-Pizarro in Fig.~\ref{fig2}, containing 169 observations in the V photometric band, show maximum brightness deviations of $\sim$2 magnitudes during its opposition in 1996 (upper plot). The lower plot demonstrates that the largest brightness deviations appear near the object's perihelion, on heliocentric distances between 2.6 and 2.8 AU, which has been suggested as an indication of activity caused by thermal processes such as sublimation of ice \citep{HandJ2006}.

The example of 133P/(7968) Elst-Pizarro demonstrates that the cometary appearance of objects can be detected photometrically by calculating deviations from their expected brightnesses. Thus, the method has potential to be applied in the future sky survey programmes. Projects like the Large Synoptic Sky Telescope (LSST) and the Panoramic Survey Telescope \& Rapid Response System (Pan-STARRS) could implement an alert system in their observing programmes, notifying when there are objects with deviations from their expected brightnesses. Unfortunately, the Gaia spacecraft (Global Astrometric Interferometer for Astrophysics) will not observe for many oppositions, but the database generated during the mission will still be suitable to search for photometric outbursts.

\begin{table}
\caption{\label{tb1}Extracted object candidates in our search for comet-like Main Belt objects and their orbital elements. a - Semi-major axis, b - Eccentricity, c - Inclination, d - Perihelion distance}
\centering
\begin{tabular}{lcccc}
\hline\hline \vspace{0.1cm}
Object & a (AU)$^{a}$ & e$^{b}$ & i ($^\circ$)$^{c}$ & q (AU)$^{d}$\\ 
\hline
(315) Constantia & 2.241 & 0.168 & 2.427 & 1.866 \\ 
(1026) Ingrid & 2.255 & 0.181 & 5.398 & 1.846 \\ 
(3646) Aduatiques & 2.755 & 0.105 & 0.589 & 2.466 \\ 
(24684) 1990 EU4 & 2.319 & 0.080 & 3.943 & 2.133 \\ 
(35101) 1991 PL16 & 2.590 & 0.180 & 12.238 & 2.124 \\ 
\hline
\end{tabular}
\end{table}

In the following sections, a short overview for each of the five candidates is given, with ideas about possible causes of the brightness deviations.

\subsection{(315) Constantia}

The asteroid (315) Constantia is a small object belonging to the Flora family in the inner Main Belt \citep{Zappala1995}. It was discovered by Johann Palisa on September 4, 1891 in Vienna. According to the JPL Small-Body Database, its absolute magnitude is H = 13.2, which corresponds to an estimated diameter of 5-12 km.

\begin{figure}
\centering
\includegraphics[width=8cm]{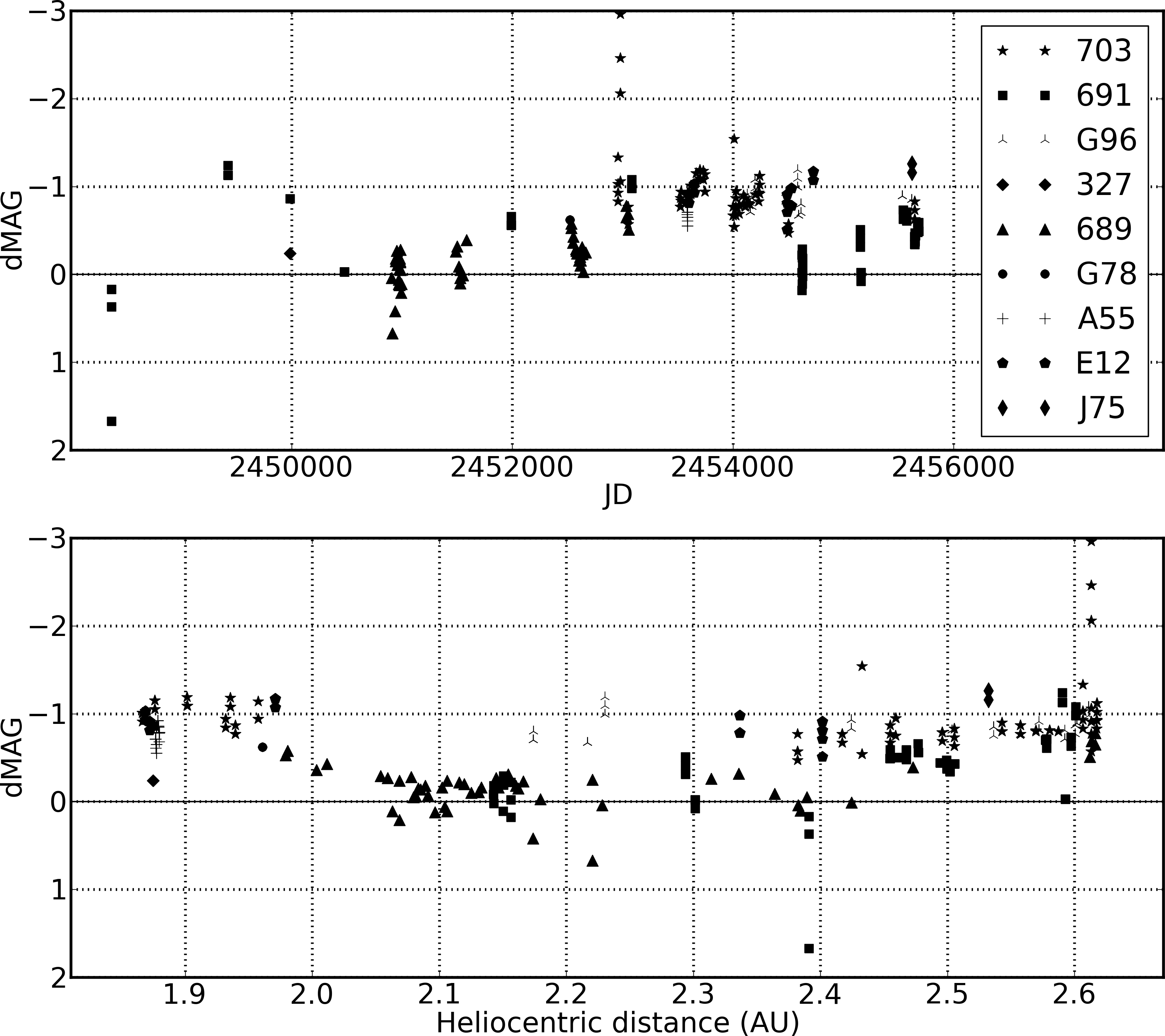}
\caption{Data plots for asteroid (315) Constantia, containing 314 points. The upper plot shows the brightness deviations versus time. The lower plot shows the brightness deviations versus heliocentric distance. The various markers represent the different observatory codes assigned by the Minor Planet Center.}
\label{fig3}
\end{figure}

The upper plot in Fig.~\ref{fig3} shows long-term brightness deviations lasting several oppositions and reaching their maximum approximately in June 2005 (JD $\sim$2453520), showing average deviations roughly estimated at 1 magnitude. The points within individual oppositions are spread over $\sim$0.5 magnitudes, which allows us to roughly estimate the object's lightcurve amplitude. Constantia rotates with a synodic period of 5.345$\pm$0.003 h showing a lightcurve amplitude of 0.57$\pm$0.2 mag, which is in good agreement with our estimation \citep{Oey2009}.

The plot shows a short brightness increase on December 05, 2003 (JD $\sim$2452979), containing only four measurements collected by the Catalina Sky Survey at the Steward Observatory (observatory code 703). Most probably it was caused by some artefacts in the images produced due to a $\sim$7 mag. bright star $\sim$35 arcmin from the object.

The lower plot in Fig.~\ref{fig3} shows no correlation between the object's heliocentric distance and its brightness deviations. This excludes thermal processes and sublimation as an explanation for its activity. If the brightness increase is caused by a physical mechanism, a possible interpretation might be connected to some long-lasting processes, for example electrostatic ejection of sub-micron grains.

The ejection of particles large enough to scatter optical photons (>0.1 $\mu$m) by the electrostatic forces is possible for asteroids up to about 10-20 km. Assuming that the grains and the asteroids are spherical, the criterion for the critical grain size for electrostatic ejection is \citep{Jewitt2012}
$$ a = \left(18 \epsilon_{0} V E l \over 4 \pi G \rho^{2} r^{2}\right) ^{1/2},  \eqno(3)$$
where
$\epsilon_{0}$ = 8.854 $\times$ 10$^{-12}$ F\,m$^{-1}$ is the permittivity of free space;
$V$ is the potential of the grains;
$E$ is the local electric field gradient;
$l$ is the shielding distance that effectively neutralizes the gradient;
$G$ is the gravitational constant;
$\rho$ is the density, assuming that the density for the object and its grains are the same, using for both 2000 kg\,m$^{-3}$; and 
$r$ represents the object's diameter.

If we assume an albedo of $\sim$0.14 for (315) Constantia, which is a typical value for the Flora family S-type asteroids \citep{Tedesco1979, Helfenstein1994}, it leads to (315) Constantia's diameter of $\sim$8 km. The model of electrostatic ejection on small bodies \citep{Jewitt2012}, gives a critical grain size of 0.27-0.86 $\mu$m for a $\sim$8 km body and by substituting lunar values, $V$ = 10 V \citep{Colwell2007}, $E$ $\sim$10 to 100 V\,m$^{-1}$, and $l$ = 1 m \citep{Colwell2007, Farrell2007}.

The Flora family is believed to originate from the catastrophic disruption of an asteroid or binary asteroid pair \citep{Tedesco1979}. A study of the dynamical dispersion of the proper eccentricity and inclination suggests that the Flora family dynamically disperses on a time scale of few 10$^{8}$ years and that its age may be significantly less than 10$^{9}$ years \citep{Nesvorn2000}. Additionally, estimates of the cratering age of (951) Gaspra's surface, which also belongs to the Flora family, suggests that the family's age is 20 to 300 million years \citep{Veverka1994}.

The fact that (315) Constantia belongs to the relatively recently created Flora family does not exclude the possibility that the activity is triggered by some other processes on its young surface, such as space weathering, which can occur on asteroid surfaces as shown by the results of the Galileo spacecraft \citep{Chapman1996}.

Another possible cause for the apparent brightness increase, not connected to any physical mechanism, may be inaccurate estimations of the predicted brightnesses. The H-G magnitude system predicts the magnitude of an object as a function of phase angle, but does not take into account the aspect angle. In order to correct the variations with regard to the aspect angle (i.e. the angle between the object's rotation axis and the observer), it is required to know the object's pole orientation.

In order to examine the object's brightnesses depending on the aspect angle $A$, we consider a three-axial ellipsoid model (asteroid) with axis $a>b>c$. The asteroid's maximum ($S_{max}$) and minimum ($S_{min}$) projected areas as seen by a distant observer, as a function of the aspect angle, are given by \citep{Pospieszalska1985}:
$$ S_{max} = \pi a b c \left({sin^{2}(A) \over b^{2}} + {cos^{2}(A) \over c^{2}} \right)^{1/2} \eqno(4)$$
$$ S_{min} = \pi a b c \left({sin^{2}(A) \over a^{2}} + {cos^{2}(A) \over c^{2}} \right)^{1/2}. \eqno(5)$$
Assuming that the projected area is proportional to the observed brightness of the asteroid, the projected area can be transformed into apparent magnitudes by $mag = -2.5 \log{S}$.

\begin{figure}
\centering
\includegraphics[width=8cm]{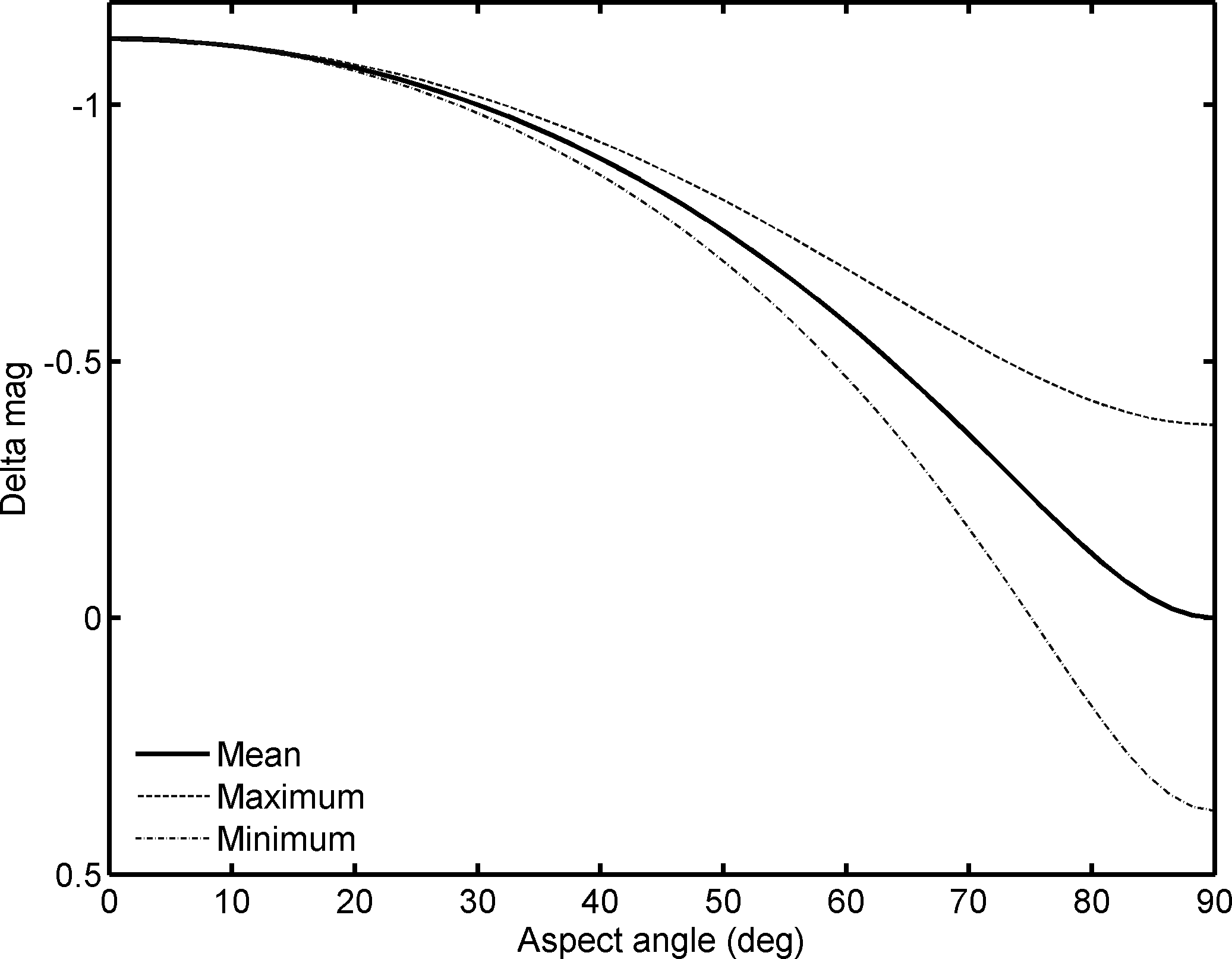}
\caption{The maximum, minimum, and mean (computed from the maximum and minimum) magnitudes are plotted as a function of aspect angle for an elongated three-axial ellipsoid with an axis ratio of 1:0.5:0.25}
\label{fig4xx}
\end{figure}

By assuming an elongated three-axial ellipsoid with an axis ratio of 1:0.5:0.25, the maximum variation in brightness due to different aspect angles (maximum difference between the maximum brightness at aspect angle 0$^{\circ}$ and the minimum brightness at aspect angle 90$^{\circ}$) would be of the order of $\sim $1.5 mag (Fig.~\ref{fig4xx}). Therefore, owing to variation in aspect angles we would expect a maximum deviation of $1.5/2 = 0.75$ magnitudes with respect to the mean. This is smaller than the typical threshold for brightness increases that we use, which is more than $\sim$1 magnitude. However, assuming that the used absolute magnitudes provided by the MPCORB are inaccurate and by accomplishing some corrections, the aspect angle variations could offer an explanation for the brightness deviations.

In addition, the object was observed from San Pedro de Atacama (observatory code I16) during two nights from January 12-14, 2013 using a 0.41 m f/3.7 telescope with a 4008 x 2672 pixel CCD camera. In total 41 images were collected, using exposure times of 240 seconds and 300 seconds in clear filter. By a visual examination of the images, no cometary features of (315) Constantia were detected. Its point spread function does not differ from stars of similar brightnesses and the measured brightness showed no deviations from the object's predicted brightness. The object's heliocentric distance at the time our observations were collected was $ \sim $2.148 AU. By looking up the given heliocentric distance in the lower plot of Fig.~\ref{fig3}, it is easy to notice that in this region brightness deviations have never been observed. Therefore, it is not surprising that the object showed no brightness deviations, but the cause of this behaviour is still speculative.

\subsection{(1026) Ingrid}

The asteroid (1026) Ingrid was discovered by Karl Wilhelm Reinmuth on August 13, 1923 in Heidelberg. It is interesting that (1026) Ingrid also belongs to the Flora family \citep{Zappala1995}, like the MBC candidate (315) Constantia, and like C/2010 A2 \citep{Snodgrass2010}. If we assume a typical albedo value for the Flora family members of $\sim$0.14, (1026) Ingrid's diameter is estimated at $\sim$13.5 km.

\begin{figure}
\centering
\includegraphics[width=8cm]{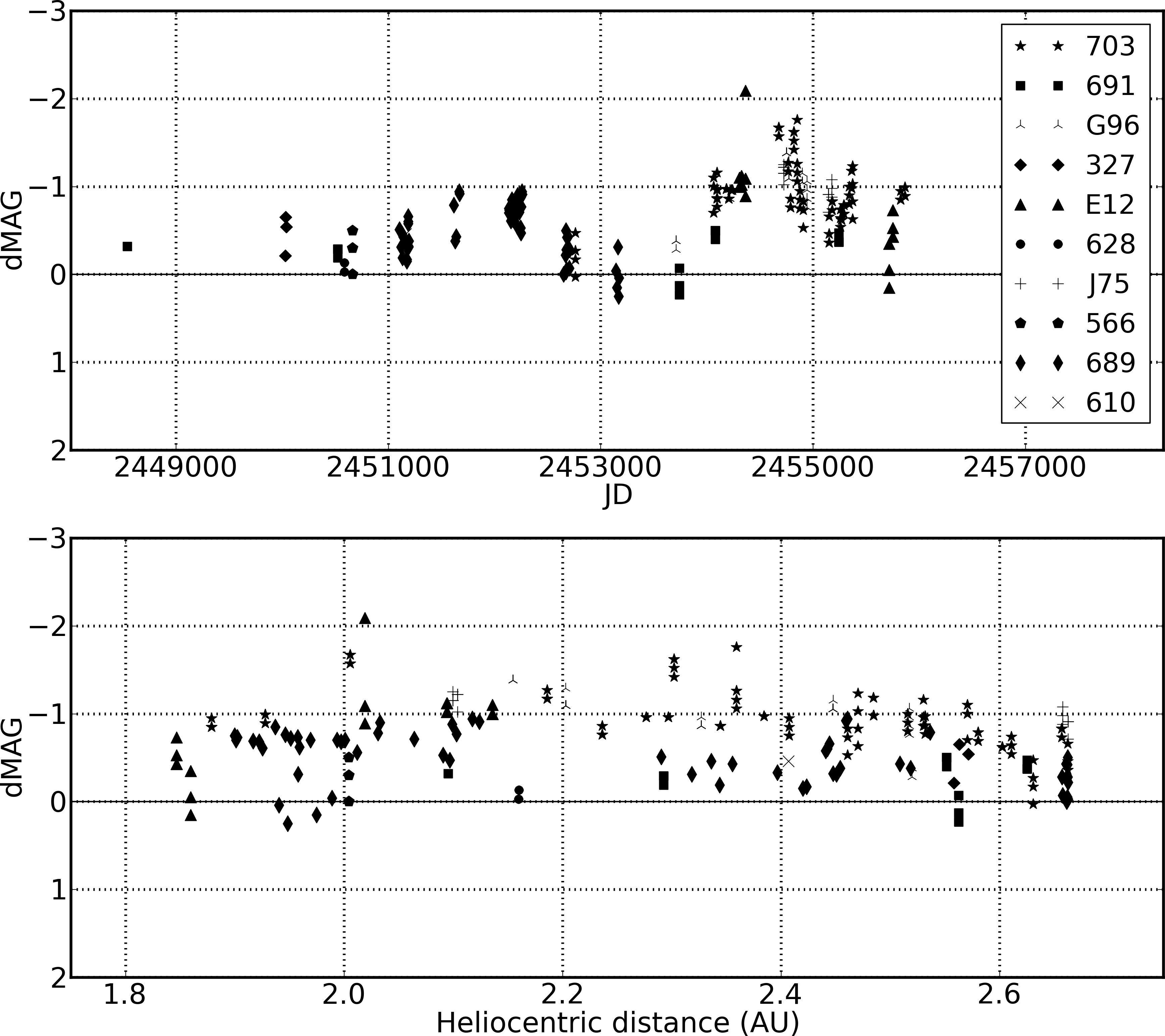}
\caption{Data plots for (1026) Ingrid containing 235 points. The upper plot shows the brightness deviations versus time. The lower plot shows the brightness deviations versus heliocentric distance. The various markers represent the different observatory codes assigned by the Minor Planet Center.}
\label{fig4}
\end{figure}

The upper plot in Fig.~\ref{fig4} shows long-term brightness deviations showing two maxima. The first maximum occurred during the oppositions in March 2000 and October 2001, showing average deviations of nearly one magnitude. The second maximum occurred during the opposition in November 2008, with maximum deviations of $\sim$1.5 magnitudes, but in this case we think that the absolute magnitude was probably estimated to be too high. A value set for $\sim$0.5 magnitudes lower would be more likely. In this case our brightness deviations would lower by the same amount.

(1026) Ingrid's rotational period of 5.3$\pm$0.3 hours was determined by \cite{Szekely2005}, showing an amplitude of $\sim$0.5 mag, which is in good agreement with the brightness deviations of observations within individual oppositions, visible in the upper plot.

The lower plot in Fig.~\ref{fig4} shows that the long-term brightness deviations are independent of the object's heliocentric distance. The origin of the deviations remains unexplained. If we consider electrostatic ejection of dust particles from the body's surface as a potential mechanism, the critical grain size for (1026) Ingrid, calculated by Eq. 3, would be 0.16-0.51 $\mu$m, which is still large enough to scatter optical photons.

The object was observed during three nights between January 14 and January 16, 2013, at the heliocentric distance of $\sim$2.645 AU, which is very near its aphelion. Using the same equipment at San Pedro de Atacama, as was used to acquire the observations of (315) Constantia, we obtained 43 images in total, using exposure times of 300 seconds, but the object does not show any unusual activity in the visual appearance, nor in its brightness.

\subsection{(3646) Aduatiques}

The asteroid (3646) Aduatiques is a 7-15 km sized Main Belt object discovered on September 11, 1985 by Henri Debehogne at La Silla. 

\begin{figure}
\centering
\includegraphics[width=8cm]{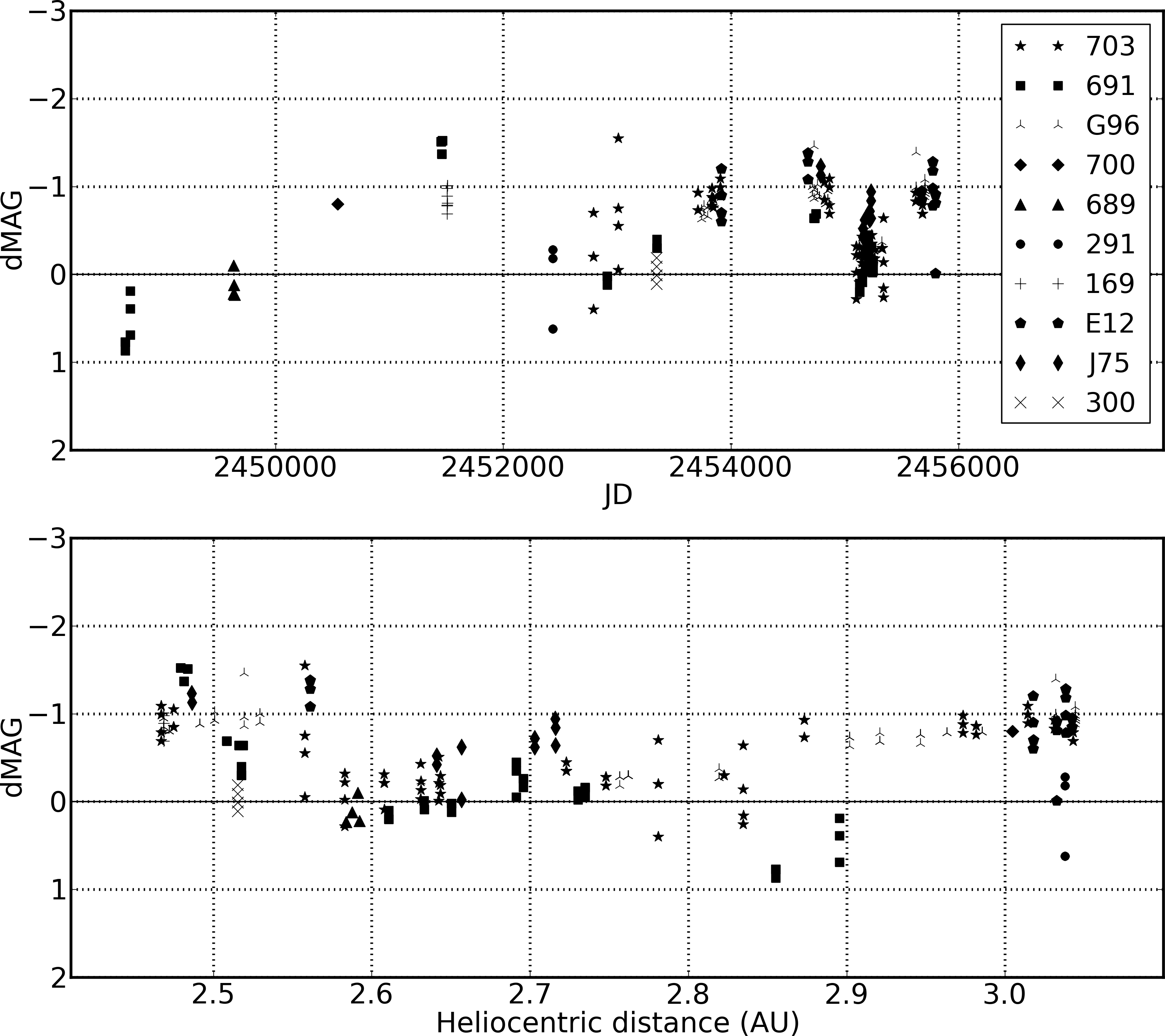}
\caption{Data plots for (3646) Aduatiques containing 260 points. The upper plot shows the brightness deviations versus time. The lower plot shows the brightness deviations versus heliocentric distance. The various markers represent the different observatory codes assigned by the Minor Planet Center.}
\label{fig5}
\end{figure}

The upper plot in Fig.~\ref{fig5} shows long-term brightness deviations in a range roughly estimated at one magnitude. The lower plot in Fig.~\ref{fig5} shows the correlation between the brightness deviations and the object's heliocentric distance. The data between $\sim$2.45 AU and $\sim$2.85 AU shows a possible correlation with the heliocentric distance, but the increasing deviations at >2.85 AU are inconsistent with this statement.

To check if the object displays cometary features like comae or tails, so far we have requested eight images from Catalina Sky Survey's Mt. Lemmon Survey (observatory code G96) which were taken at the moment of the largest deviations. The requested images containing (3646) Aduatiques are listed in Table~\ref{tb2}.

\begin{table}
\caption{\label{tb2} Requested data of (3646) Aduatiques from the Mt. Lemmon Survey and its brightness measurements compared to the expected brightness. a - Brightness reported by the Mt. Lemmon Survey, taken from the MPCAT-OBS Observation Archive, b - Remeasured brightness using the requested images, c - Expected brightness generated by the JPL Solar System Dynamics database, d - Difference between the remeasured and expected brightness}
\centering
\begin{tabular}{lcccc}
\hline\hline \vspace{0.1cm}
 Date&Obs.$^{a}$ &Remeas.$^{b}$ & Exp.$^{c}$ &$\Delta$mag$^{d}$\\
 & mag & mag & mag\\
\hline
2008 09 20.35225 & 15.9 & 16.48 & 17.38 & -0.90  \\  
2008 09 20.36087 & 16.4 & 16.46 & 17.38 & -0.92  \\
2008 09 20.36956 & 16.5 & 16.47 & 17.38 & -0.91  \\
2008 09 20.37825 & 16.4 & 16.43 & 17.38 & -0.95  \\ 
2011 05 25.30260 & 17.2 & 17.29 & 18.14 & -0.85  \\ 
2011 05 25.31301 & 17.2 & 17.28 & 18.14 & -0.86  \\ 
2011 05 25.32345 & 17.2 & 17.27 & 18.14 & -0.87  \\ 
2011 05 25.33391 & 17.2 & 17.28 & 18.14 & -0.86  \\ 
\hline
\end{tabular}
\end{table}

The images were obtained in Mt. Lemmon's regular sky survey programme, using their 1.5 m f/2.0 Cassegrain reflector. The images were exposed 30 (images taken in 2008) and 40 seconds (images taken in 2011), which, for the observed object, contributed to a signal-to-noise ratio of around 60. By visual examination of the images, no typical indications of cometary activity was found, besides the negative deviation in (3646) Aduatiques' brightness.
To re-analyse the object's brightness, the images were remeasured using a Windows version of SExtractor and compared to the expected brightness (Table~\ref{tb2}) generated by the JPL Solar System Dynamics database \citep{Chamberlin1997}.

At the moment, it is difficult to guess what mechanisms cause the large variations in (3646) Aduatiques, and future observations and a rotational lightcurve will be required. The upper plot allows us to estimate the object's lightcurve amplitude of $\sim$0.6 magnitudes.

\subsection{(24684) 1990 EU4}

The asteroid (24684) 1990 EU4 is a $\sim$3-6 km Main Belt object, discovered by Eric Walter Elst on March 02, 1990 from La Silla.

\begin{figure}
\centering
\includegraphics[width=8cm]{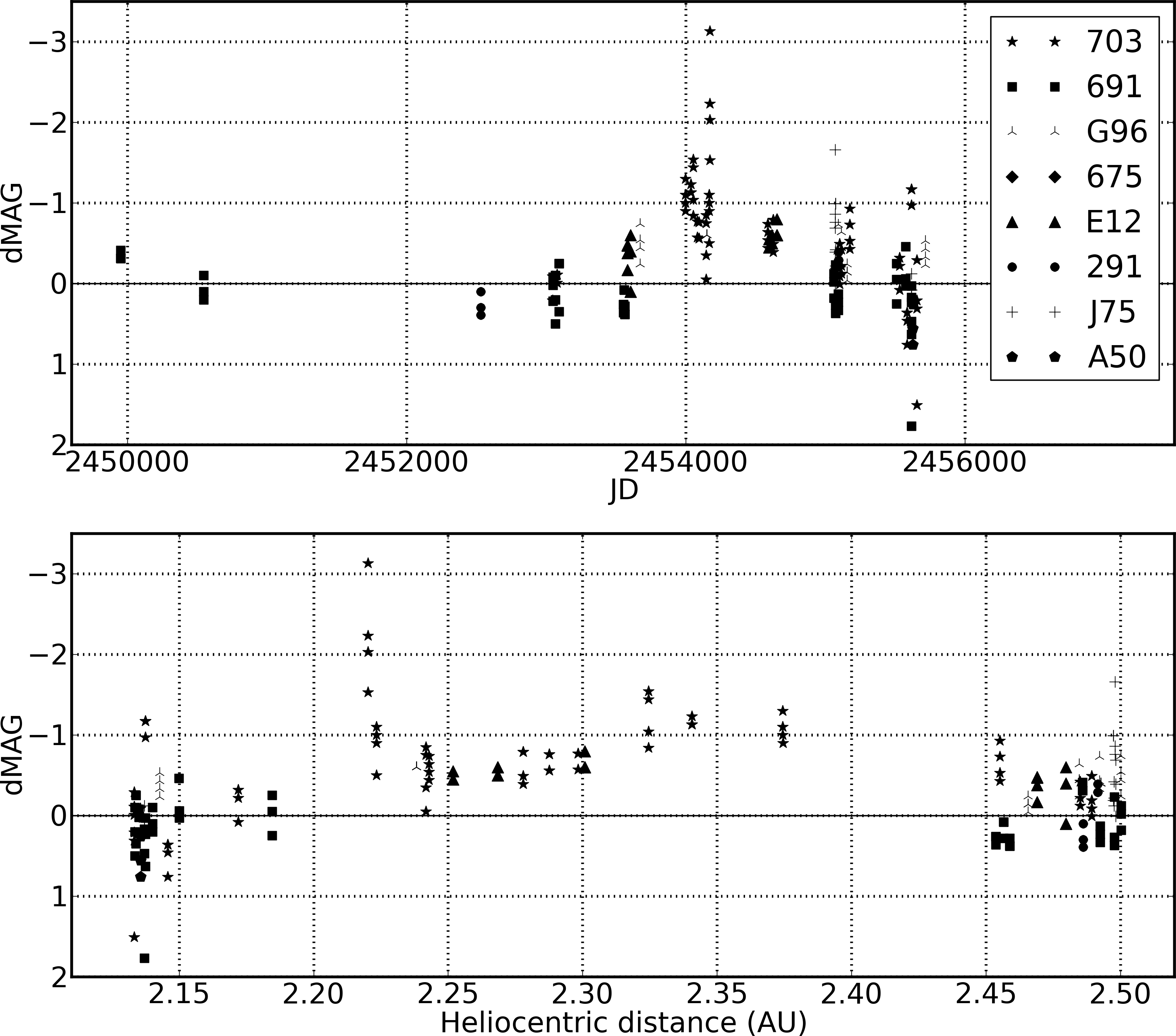}
\caption{Data plots for (24684) 1990 EU4 containing 190 points. The upper plot shows the brightness deviations versus time. The lower plot shows the brightness deviations versus heliocentric distance. The various markers represent the different observatory codes assigned by the Minor Planet Center.}
\label{fig6}
\end{figure}

The plots in Fig.~\ref{fig6} show long-term brightness deviations of nearly one magnitude. The object's rotational lightcurve is still unknown, but according to the upper plot in Fig.~\ref{fig6}, we think that it may have an amplitude of $\sim$0.5 magnitudes because this is the typical scatter in single oppositions. The brightness increase visible in the upper plot reaching brightness deviations of $\sim$3 magnitudes was observed on March 13, 2007 (JD ~2457825), by the Catalina Sky Survey (observatory code 703) and was most probably caused by artefacts in the images produced by bright stars in the object's surrounding.
 
The lower plot in Fig.~\ref{fig6} shows no correlation of the brightness deviations versus heliocentric distance, which could indicate mechanisms similar to those causing deviations on (315) Constantia and (1026) Ingrid.

\subsection{(35101) 1991 PL16}

The asteroid (35101) 1991 PL16 was discovered at the Palomar Observatory on August 07, 1991, by Henry E. Holt. The 5-12 km sized object belongs to the Eunomia family \citep{Zappala1995}, and was possibly formed from the fragmentation of a partially differentiated S-type parent body. The Eunomia family's age is suggested to be similar to the Flora family's age \citep{Lazzaro1999}.

\begin{figure}
\centering
\includegraphics[width=8cm]{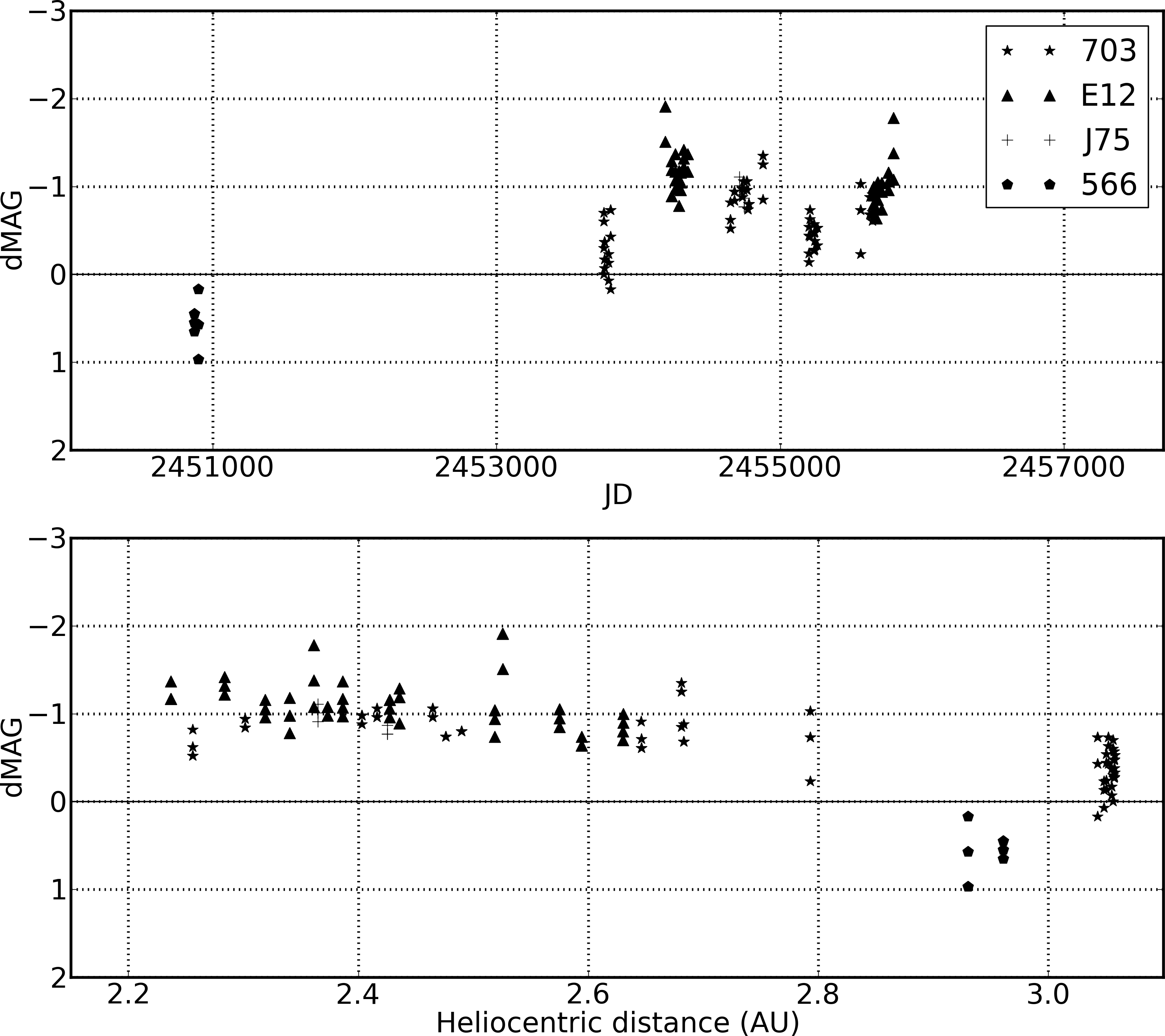}
\caption{Data plots for (35101) 1991 PL16 containing 143 points. The upper plot shows the brightness deviations versus time. The lower plot shows the brightness deviations versus heliocentric distance. The various markers represent different observatory codes assigned by the Minor Planet Center.}
\label{fig7}
\end{figure}

The upper plot in Fig.~\ref{fig7} shows long-term average deviations of maximum one magnitude. The object's rotational period is still unknown, but according to the deviations in individual oppositions, we expect its amplitude to be slightly above $\sim$0.5 magnitudes.
The lower plot shows a possible correlation of its brightness deviations versus the heliocentric distance, which could indicate a thermal process such as sublimation of ice.

One more mass-loss mechanism that may be responsible for the activity is rotational instability. \cite{Jewitt2012} offered a model to estimate the objects' critical rotational period $P_{c}$, that depends on its shape and density, given by
$$ P_{c} = k \left[3 \pi \over G \rho\right] ^{1/2},  \eqno(4)$$
where
$k$ is the axial ratio of the body, given as $k = a/b$, if axis a $\geqslant$ b;
$G$ is the gravitational constant; and
$\rho$ is the density for the object.

Assuming that the bulk density of all Eunomia family members is similar to their parent body's density, and using a lower estimations for (15) Eunomia's density of 960 $\pm$ 300 kg\,m$^{-3}$ \citep{Hilton1997, Tedesco1992}, assuming a prolate body with $k$ = 2, Eq. 4 gives a value for the critical rotational period of $\sim$6.74 h. Therefore, if the rotational period of (35101) 1991 PL16 is shorter than 6.74 h, rotational instability might be a possible explanation for the object's activity.

For comparison, the average bulk density of S-type asteroids is 2720 $\pm$ 90 kg\,m$^{-3}$ \citep{Britt2002, Carbognani2011}. By choosing axial ratios $k$ between 1 and 2, the estimated critical rotational period is between $\sim$2 and $\sim$4 hours. This excludes rotational instability as a source mechanism for the activity of our candidates (315) Constantia and (1026) Ingrid, because their rotational periods are greater than 4 hours.

\section{Conclusions}

Our photometric search for active Main Belt objects was carried out just by using the MPCAT-OBS Observation Archive. From $\sim$75 million observations in total, covering $\sim$300\,000 numbered objects, we extracted five new candidates of photometrically active Main Belt objects and the already known Main Belt comet 133P/(7968) Elst-Pizarro. The detection of 133P/(7968) Elst-Pizarro encourages us that our method can give positive results in searching for objects with signs of activity. Other already known Main Belt comets have not been detected using our method because they are still unnumbered, and are therefore not included in the used MPCAT-OBS data set, or their number of observations was still too low to fulfil our filtering requirements. Consequently, this means that the detected candidates cannot be used to make estimations of the statistical evidence of the quantity of similar objects in the Main Belt.

We believe that the possible activities on our candidates have remained unnoticed until now because they have been evident only in the objects' brightness deviations, instead of showing typical cometary signs like comae or tails.

An examination of images of (3646) Aduatiques, provided by the Mt. Lemmon Survey, shows no visible tail or coma at the moment of the largest deviations, but has proven the brightness deviations.

The objects (315) Constantia, (1026) Ingrid, (3646) Aduatiques, and (24684) 1990 EU4 show brightness deviations independent of their heliocentric distances. For these objects, electrostatic ejection of dust grains or space weathering processes could offer possible explanations for the activities. Until now, there have been no clear examples of objects whose activities are caused by these processes. On the other hand, we cannot exclude that some brightness deviations are caused by some other, non-physical mechanism. By modelling the objects' brightnesses depending on the aspect angle, we have shown that for highly elongated objects deviations of $ \sim $0.75 magnitudes are likely, and could offer possible explanations of the observed brightness deviations.

An interesting fact is that three of our new candidates belong to relatively similar Main Belt families - (315) Constantia and (1026) Ingrid to the Flora family, and (35101) 1991 PL16 to the Eunomia family. Because of their young surface materials, this might be one of the main connections to the possible activities.
By assuming an average value for the objects' bulk densities, rotational instability as a mass-loss mechanism on (315) Constantia and (1026) Ingrid can be excluded. The object (35101) 1991 PL16 shows brightness deviations dependent on its heliocentric distance, which could indicate a thermal process responsible for the activity.
An additional possible explanation for weak activities on small bodies is seismic shaking, induced by the release of energy by the liberation of internal stresses, thermal cracking, or small impacts.

One of the disadvantages of our method is the relatively low accuracy of photometric data contained in the MPCAT-OBS Observation Archive, which is a result of many different instrumental set-ups, but the method could be successfully implemented in future sky surveys that assure frequent imaging of the same objects with good photometric accuracy. 
These very simple alert systems could enable us to detect outbursts on Main Belt objects almost simultaneously with their occurrence, which is potentially interesting when searching for impact events in the Main Belt. 

The method can be used to reveal objects with outbursts, but also objects with peculiar photometric behaviour, signifying very elongated objects or binary objects that possess large rotational lightcurve amplitudes, which can be interesting side products of the search. Hence, further observations of the object candidates should be made in the future to determine whether these objects are really active, and if so, to investigate the cause of the activity.

\begin{acknowledgements}
This  work  was  supported in part by the Catalina Sky Survey’s Mt. Lemmon Survey (observatory code G96), which provided us with the requested images of (3646) Aduatiques. Support from AYA2011-30106-C02-01 and FEDER funds is acknowledged.
\end{acknowledgements}

\end{document}